# Sensitivity of Exciton Spin Relaxation in Quantum Dots to Confining Potential

S. Mackowski*, T. Gurung, H. E. Jackson, and L. M. Smith

*Department of Physics, University of Cincinnati, 45221-0011 Cincinnati, OH*

W. Heiss

*Institut für Festkörper- und Halbleiterphysik, Universität Linz, 4040 Linz, Austria*

J. Kossut and G. Karczewski

*Institute of Physics Polish Academy of Sciences, Lotnikow 32/46, 02-668 Warsaw, Poland*

We observe a strong dependence of the exciton spin relaxation in CdTe quantum dots on the average dot size and the depth of the confining potential. For the excitons confined to the as-grown CdTe quantum dots we find the spin relaxation time to be 4.8 ns. After rapid thermal annealing, which increases the average dot size and leads to weaker confinement, we measure the spin relaxation tine to be 1.5 ns, resulting in smaller values of the absolute polarization of the quantum dot emission. This dramatic enhancement of the spin scattering efficiency upon annealing is attributed to increased mixing between different spin states in larger CdTe quantum dots.

*author to whom the correspondence should be addressed: electronic mail: seb@physics.uc.edu

The manipulation of the exciton spin properties in semiconductor nanostructures requires precise control of the exciton energy levels. This may be achieved either by careful design of the structure or by applying electromagnetic fields. In the case of semiconductor quantum dots (QDs) the internal exciton structure is very sensitive to the size and shape of the QD [1]. Depending on the symmetry of the QD potential the ground state of the exciton is either twofold degenerate (symmetric QD) or split by the exchange interaction (asymmetric QD) [2]. Moreover, energy levels of excitons in QDs are also affected by external magnetic fields via the Zeeman interaction [2,3], which leads to spin splitting determined by the effective g-factor. A number of studies have shown that the degeneracy of the exciton levels significantly impacts the exciton spin relaxation in semiconductor QDs. For the split energy levels (either by asymmetry [4-6] or by magnetic field [6-8]) the spin of the excitons is conserved throughout their lifetime. In contrast, when the exciton levels are degenerate, the exciton spin relaxation is usually extremely rapid [7].

Recent theoretical models have determined the admixture of exciton states in QDs, mainly due to spin – orbit and the electron – hole exchange interactions, as the major mechanisms of the exciton spin relaxation in QDs [9-11]. Thus, the exciton spin relaxation in QDs also depends on the energy separation between the ground states and the excited states (such as light-hole exciton levels). Clearly, the separation between heavy- and light-hole energy levels decreases for larger QD sizes [12], which increases the mixing between states with different spin [11]. As a result, exciton spin relaxation in QDs should strongly depend not only on the symmetry of the QD potential or external magnetic fields but also on the QD size [11].

In this letter, we demonstrate a way of tuning the exciton spin relaxation time in QDs through rapid thermal annealing. Using a recently developed method [13] to analyze the polarization of the PL emission for large QD ensembles, we find that the exciton spin relaxation time of the annealed CdTe QDs is reduced by over a factor of two compared to the as grown QD sample. As reported previously, annealing of CdTe QDs results in larger QDs with shallower confining potentials caused by interdiffusion of Zn and Cd [14]. Therefore, we attribute the change in the spin dynamics of the excitons in the annealed QDs to stronger mixing between different spin states in larger CdTe QDs.

The sample containing CdTe QDs was grown by molecular beam epitaxy on a (100) - oriented GaAs substrate. The dots formed by depositing 4 monolayers of CdTe on top of 2 μm – thick ZnTe buffer were capped by a 50 nm - thick ZnTe layer. In order to change the QD morphology the sample was annealed at a temperature $450^0$C in an argon atmosphere for 15 seconds. A detailed description of the annealing-induced changes can be found elsewhere [14].

The spin relaxation of excitons confined to QDs was examined by resonantly excited polarized PL spectroscopy. Circularly polarized excitons were excited with an $Ar^+$ laser-pumped Pyrromethane dye laser. The measurements were performed in a continuous-flow helium cryostat with magnetic fields up to 4 T applied perpendicular to the dot layer. Two sets of Babinet – Soleil compensators combined with Glan – Thompson linear polarizers were used to precisely control the polarization of the excitation, as well as to analyze the polarization of the QD emission. The signal was dispersed by a DILOR triple monochromator and then detected by a nitrogen - cooled CCD detector.

Fig. 1 (a) shows low temperature (T=5K) resonantly excited PL spectra obtained for the annealed CdTe QDs at zero magnetic field. The resonantly excited PL spectra are composed of three strong LO phonon replicas superimposed on a broad PL band. These LO phonon replicas are the spectral signatures of exciton recombination from QDs ground states populated directly through LO phonon-assisted absorption, while the broad emission is due to excited state – ground state excitations in the QD ensemble [15]. In Fig. 1 (a) both $\sigma^+$ (solid points) and $\sigma^-$ (open points) - polarized emissions measured for $\sigma^-$ - polarized excitation are displayed. It is important to emphasize that with circularly polarized excitation and detection at zero magnetic field we are only sensitive to symmetric QDs that are characterized by twofold degenerate exciton levels [2]. As can be seen, at B=0T the intensity and the shape of both $\sigma^+$ and $\sigma^-$ - polarized PL signals are identical indicating no net polarization of the QD emission. A similar result is also observed for $\sigma^+$ - polarized excitation (not shown). This result implies the spin relaxation time of the excitons confined to the symmetric QDs to be much shorter than the exciton recombination time [7]. Since the latter in the annealed QDs is 0.2 ns [14], the resulting spin relaxation time of the excitons at B=0T must be around 10ps.

The spin dynamics of QD excitons changes strongly in applied magnetic fields. In Fig. 1 (b) we show resonantly excited PL spectra measured at B=3T for $\sigma^-$ polarized excitation. Solid and open symbols represent $\sigma^+$ and $\sigma^-$ - polarized emissions, respectively. In contrast to the zero field spectra, the QDs emission in external magnetic fields is polarized in the same orientation as the excitation [7,8]. The strong PL polarization means that the excitons excited using LO phonon-assisted absorption, maintain their spin polarization within their lifetime. We attribute this dramatic increase of the exciton spin relaxation time to the Zeeman interaction, which lifts the degeneracy of spin-polarized exciton energy levels in QDs [2]. Previous measurements of the as-grown CdTe QDs have shown an increase of the exciton spin relaxation time whenever the exciton states in QDs are non-degenerate [7]: either due to deformation of the confining potential, or the application of external magnetic field. We note, however, that there are important differences between the annealed QDs and these as-grown QDs, which we shall discuss in detail below.

In order to quantitatively describe the dependence of the polarization on the magnetic field we fit the resonantly excited spectra by a sum of four Gaussian lineshapes: three of them being the LO phonon replicas and the fourth representing emission from excitons populated through excited state – ground state relaxation. According to the arguments described in [15], we assume that the broad emission can be modeled with the same lineshape as the non-resonant PL spectrum of the QD ensemble. In this way the intensities of the LO phonon replicas for both circularly polarized excitations were obtained as a function of magnetic field. An example of such a fit, which corresponds reasonably well to the experimentally measured spectrum, is shown in Fig. 1 (c). Further, we define the polarization $P=(I^+ - I^-)/(I^+ + I^-)$, where $I^+$ and $I^-$ correspond to the intensity of $\sigma^+$ and $\sigma^-$ - polarized LO phonon replica, respectively. In Fig. 2 we plot the absolute value of polarizations P obtained for the first LO phonon replica of the (a) as-grown and (b) annealed CdTe QDs for $\sigma^+$ (solid squares) and $\sigma^-$ (open squares) - polarized excitations. An important observation can be made: Although both samples exhibit larger PL polarization when the QDs are excited with $\sigma^-$ polarized laser, the absolute values of the polarization measured for the as-grown QDs are much larger than for the annealed QDs. Qualitatively this suggests that the exciton spin relaxation time in the annealed QDs should be shorter than in the as-grown QD sample.

The larger absolute polarization observed for $\sigma^-$ polarized excitation can be understood when taking into account the sign of the effective exciton g-factor, which has been found to be similar for both as-grown and annealed CdTe QDs and equal to –3 [16]. The negative g-factor implies that the low-energy state of the exciton in CdTe QDs is always $\sigma^-$ polarized. This leads to the asymmetry of the polarizations, as the excitons more effectively lower their energy through a spin-flip process, than the reverse. Importantly, this asymmetry of the polarization makes it possible to estimate the exciton spin relaxation time on the QDs. Following the model presented in [13], the difference between polarizations ($\Delta$P) is determined by a simple relation between the exciton spin relaxation time ($\tau_S$) and the exciton recombination time ($\tau_R$):

$$\Delta P = \frac{2\tau_R (e^{\Delta E / k_B T} - 1)}{\tau_R + e^{\Delta E / k_B T} (\tau_R + \tau_S)}$$

In this equation $\Delta$E is the exciton Zeeman splitting and T is the temperature. Using this formula the exciton spin relaxation time of 4.8 ns has been previously obtained for the as-grown CdTe QDs [13].

In Fig. 3 we plot the difference between polarizations $\Delta$P for the annealed QDs (squares) and the as-grown QDs (circles) as a function of $\Delta E/k_B T$ together with the result of the fit. The values of the effective exciton g-factor and the temperature are –3 and 5K, respectively. The experimental data were measured for the first LO phonon replicas. From this analysis we obtain the exciton spin relaxation time in the annealed CdTe QDs of 1.5 ns. We recall that for the as-grown QD sample $\tau_S$=4.8 ns has been deducted [13]. The reduction of the exciton spin relaxation time upon annealing is consistent with the lower absolute polarizations measured for the annealed QDs as compared with the as-grown QDs (see Fig. 2). We note, that the observation of smaller polarizations in the case of the annealed QDs provides an independent verification of the shorter spin relaxation time, as the absolute values of polarization are absent in Eq. 1.

Qualitative understanding of the strong reduction of the spin relaxation of the excitons confined to the annealed QDs is based on the observation that the QD emission shifts toward higher energies upon annealing [14]. At the same time, however, the observed decrease of the exciton recombination time shows that annealing leads also to an increase of the average QD size [14]. Through interdiffusion-induced partial smearing out of the barriers, the confinement of the excitons in annealed CdTe QDs weakens, resulting in a smaller average energy separation between the states with different spin. The decrease of this splitting increases the mixing between the levels, and hence, enhances the efficiency of the spin-flip transitions between the exciton levels in a QD split by the Zeeman interaction [9]. We believe that this effect is predominantly responsible for the strong reduction of the exciton spin relaxation time in the annealed CdTe QD. One could also speculate that the chemical alloying of the QDs leads to much broader energy dispersion of the acoustic phonons, that dominate the spin-flip processes in QDs [9]. These experimental studies on the impact of the QD confinement on the exciton spin dynamics could be a valuable input into the ongoing efforts to understand and model the spin properties of semiconductor nanostructures.

In summary, we observe a strong effect of the QD confining potential on the exciton spin relaxation in CdTe QDs. We find that while the absolute PL polarization measured for the annealed (larger and shallower) QDs is smaller than for the as-grown (smaller and deeper) QDs, the polarization difference between two circularly polarized excitations is much larger. Analysis of the data show that for the annealed CdTe QDs the exciton spin relaxation time is a strongly reduced to 1.5ns, over a factor of two smaller than the 4.8 ns value measured for the as-grown QDs. We attribute this strong reduction of the spin relaxation time to enhancement of the admixture between electronic levels in the larger and shallower annealed CdTe QDs.

The work was supported by NSF grants nr 9975655 and 0071797 (United States) and ÖAD project Nr. 05/2001 (Austria).

**Figure captions:**

**Figure 1.** Resonantly excited PL spectra measured at T=5K for annealed CdTe QDs at (a) B=0T and (b) B=3T. The excitation was $\sigma^-$ - polarized and both $\sigma^+$ (solid points) and $\sigma^-$ (open points) - polarized emissions were analyzed. (c) Comparison between experimentally measured PL (open points) and the fitted PL spectrum (solid line). Dashed lines represent all the components of the PL spectrum.

**Figure 2.** Polarization of the PL emission excited with $\sigma^+$ (solid symbols) and $\sigma^-$ (open symbols) polarized excitation measured for the first LO phonon replica for the (a) as-grown CdTe QDs and (b) annealed CdTe QDs.

**Figure 3.** Measured (points) and fitted (lines) values of the difference between polarizations ($\Delta P$) plotted as a function of $\Delta E/k_BT$. Squares and circles represent the results of the annealed CdTe QDs and the as-grown CdTe QDs, respectively.

Figure 1.

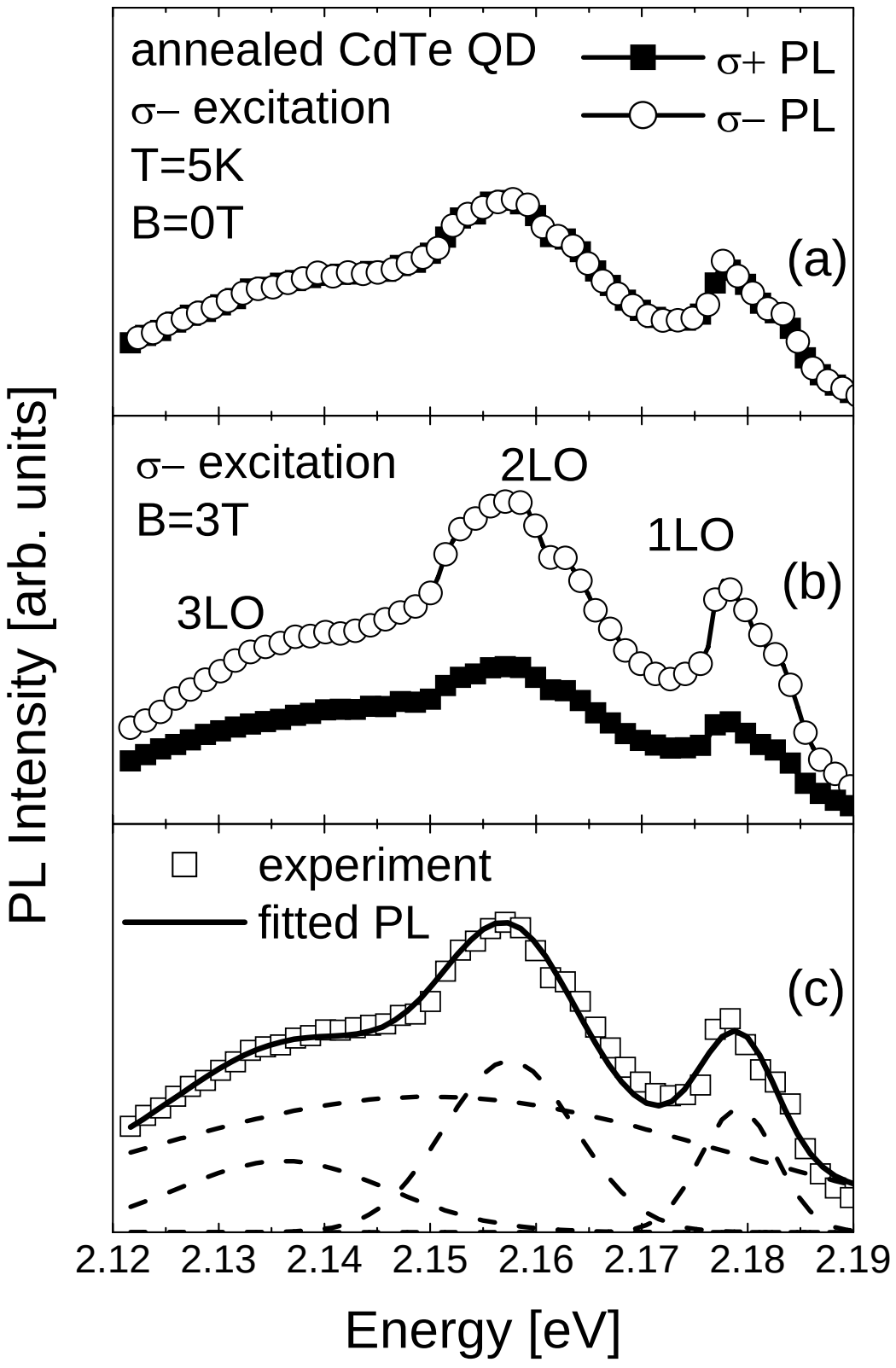
annealed CdTe QD
σ− excitation
T=5K
B=0T
σ+ PL
σ− PL
(a)
σ− excitation
B=3T
2LO
1LO
3LO
(b)
experiment
fitted PL
(c)
PL Intensity [arb. units]
2.12 2.13 2.14 2.15 2.16 2.17 2.18 2.19
Energy [eV]

Figure 2

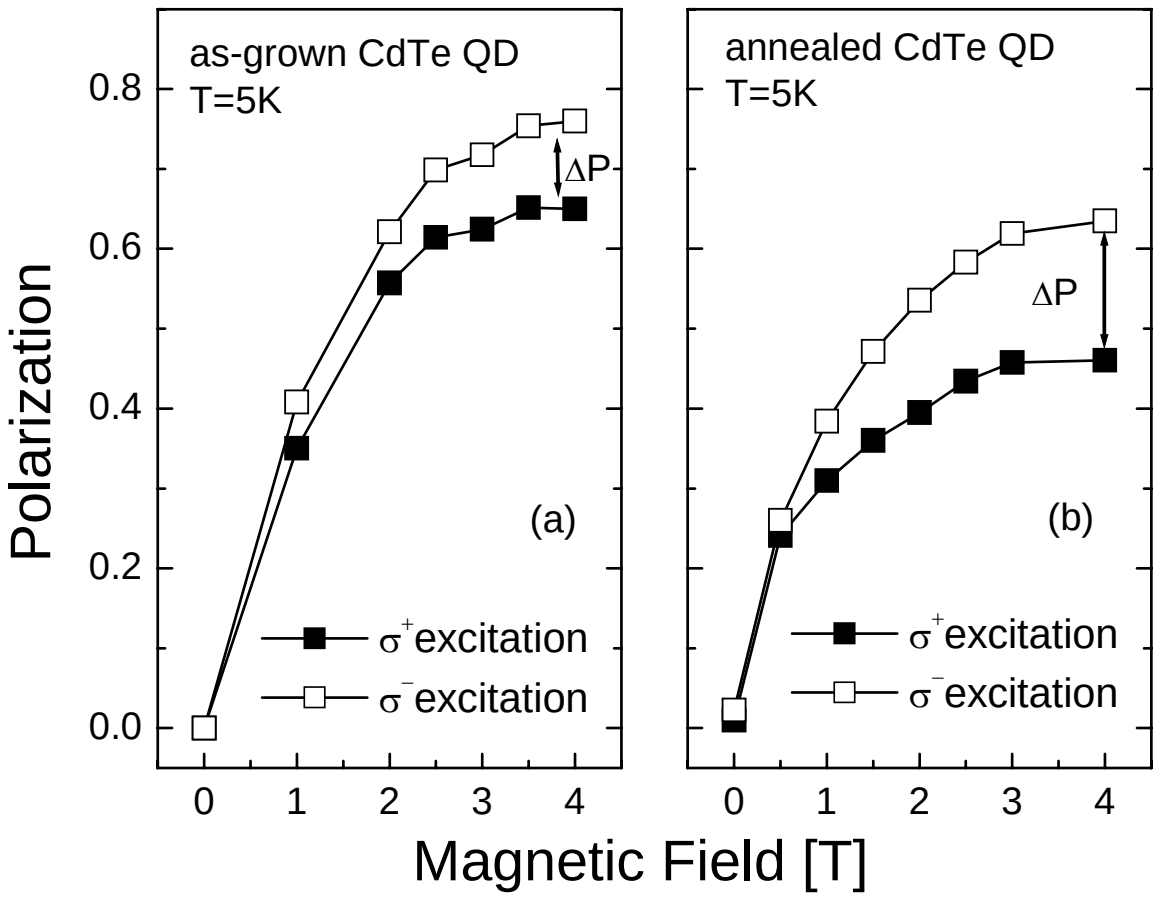

Figure 3.

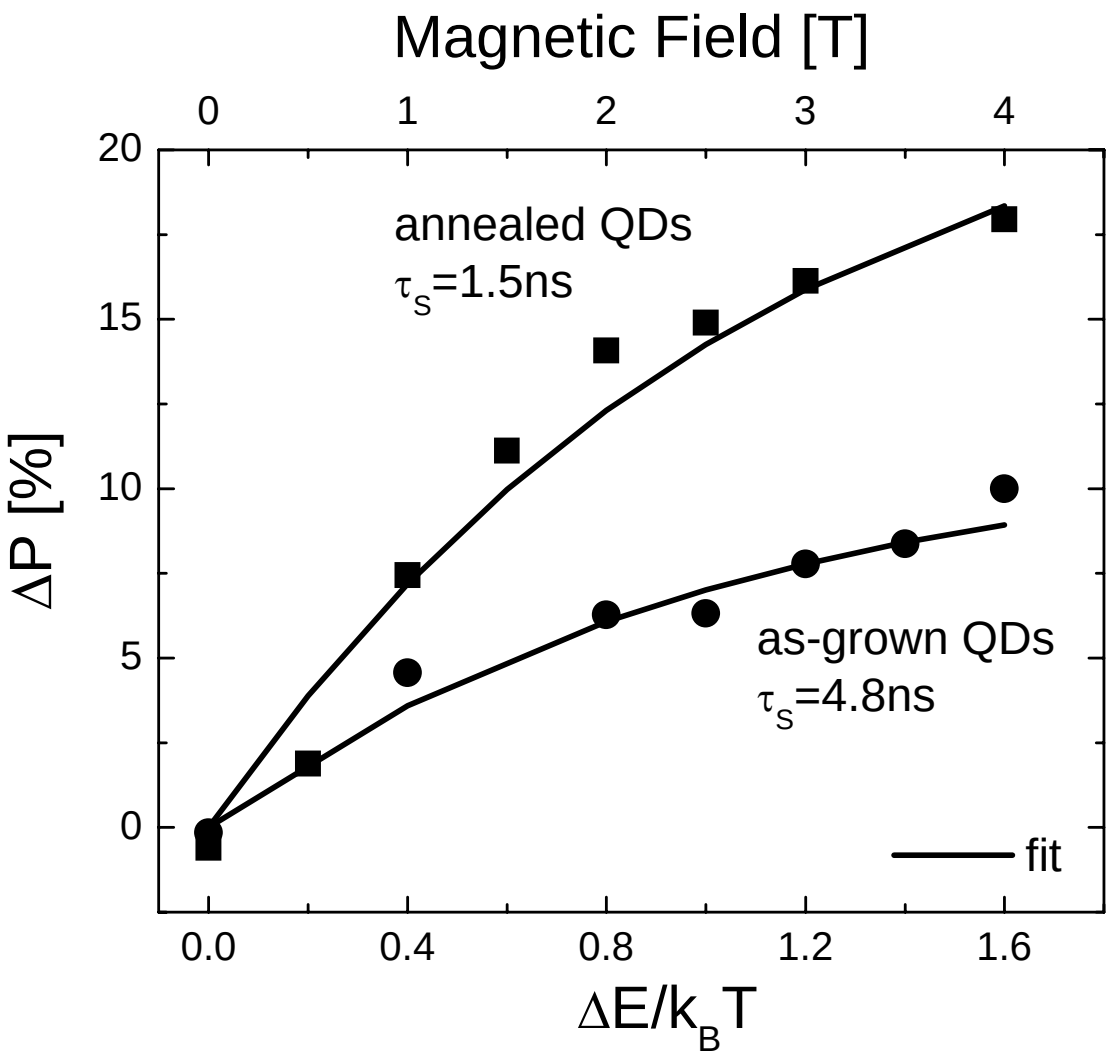